\documentclass[12pt]{article}
\usepackage{graphicx}

\begin{document}

\centerline{\bf \large Monte Carlo simulation of Ising model}
\bigskip

\centerline{\bf \large  on directed Barabasi-Albert network}

\bigskip
Muneer A. Sumour, Physics Department, Al-Aqsa University,  P.O.B. 4051, Gaza,  
Gaza Strip, Palestinian Authority

M.M. Shabat, Physics Department, Islamic University, P.O.B. 108, Gaza,
Gaza Strip, Palestinian Authority

Email:  Shabat@mail.iugaza.edu, msumoor@yahoo.com

\bigskip
Abstract:

The existence of spontaneous magnetization of Ising spins  on directed 
Barabasi-Albert networks is investigated  with seven neighbors, by using
Monte Carlo simulations. In large systems we see the magnetization 
for different temperatures $T$ to decay after a characteristic time $\tau(T)$, 
which is extrapolated to diverge at zero temperature.


\bigskip
{\bf Introduction}: 

The Ising magnet is since decades a standard tool of computational physics [1].
We apply it here to scale-free networks [2], where previous simulations  [3]
indicated a Curie temperature increasing logarithmically with increasing system
size $N$. In contrast to that work we use here directed [4] as opposed to 
undirected networks and then apply the standard Glauber kinetic Ising model [1]
to the fixed network.

\bigskip
{\bf Directed Barabasi-Albert network}:

Putting Ising spins onto the sites (vertices, nodes) of a network, we  
simulate our Ising magnetic model on directed Barabasi-Albert networks. 
The Barabasi-Albert network is grown such that the probability of a new site
to be connected to one of the already existing sites is proportional to the 
number of previous connections to this already existing site: The rich get 
richer. In this way each new site selects exactly $m$ old sites as neighbours.

Then each spin is influenced by the fixed number $m$ of neighbours which it had 
selected when joining the network. It is not influenced by other spins which 
selected it as neighbour after it joined the network.

The Barabasi-Albert network is simulated by a Fortran program calculating the 
neighbours:     

\begin{verbatim}
      parameter( nsites=500000,m=7,iseed=3, maxmax=20000,
    1 max=nsites+m, length=1+2*m*nsites+2*m*m , T=1.0)
      integer*8 ibm, iex
      dimension list(length), is(max), iex(2*m+1), neighb(max,m)
      ibm=iseed-1
      factor=(0.25d0/2147483648.0d0)/21474836484.0d0
      do 7 i=1,m
        do 7 nn=1,m
          neighb(i,nn)=nn
 7        list((i-1)*m+nn)=nn
      L=m*m
c     All m initial sites are connected
      do 1 i=m+1,max
        do 2 new=1,m
 4        ibm=ibm*16807 
          j=1+(ibm*factor+0.5)*L
          if(j.le.0.or.j.gt.L) goto 4
          j=list(j)
          list(L+new)=j
          list(L+m+new)=i
 2        neighb(i,new)=j
 1      L=L+2*m
c     print *,ibm,neigh
c     end of network and neighbourhood construction
\end{verbatim}
 
At each step, a new spin is added which builds $m$ new connections {\tt neighb},
randomly to already existing spins. The probability for an existing spin to be 
chosen as neighbour is proportional to the number of its neighbours, with the
help of the Kertesz {\tt list}. 

\bigskip
{\bf Ising Magnet using Monte Carlo Simulations}:

First we initialize a directed Barabasi-Albert network with $m$ neighbours (all 
$m$ initial spins are connected with each other and themselves), here $m=2$ and
7. 
We put an Ising spins onto every site, with all spins up, because we test here 
for ferromagnetism. Then with the standard Glauber (heat bath) Monte Carlo 
algorithm spins we search for thermal equilibrium 
at positive temperature. 

After putting all spins on the network, we go through the whole network and use 
the Monte Carlo step (MCS) on every spin; we say that we make one MCS per spin 
at each time step. Each spin is influenced by its exactly $m$ neighbours.
We calculate the magnetization versus the number of time steps, 
with the same number of neighbours $m$ and different temperatures $T$.

Initially we start with all spins up, a number of spins equal to 500,000, 
and time up to 20,000. Then we vary the temperatures and study $m=2$ and 7 
for nine samples (nine different random number sequences). The temperature is
measured in units of the critical temperature of the square-lattice Ising
model.

So we can draw a graph of magnetization versus time for different 
temperatures to see how the magnetization changes, Fig. 1. 

\begin{figure}[hbt]
\begin{center}
\includegraphics[angle=-90,scale=0.5]{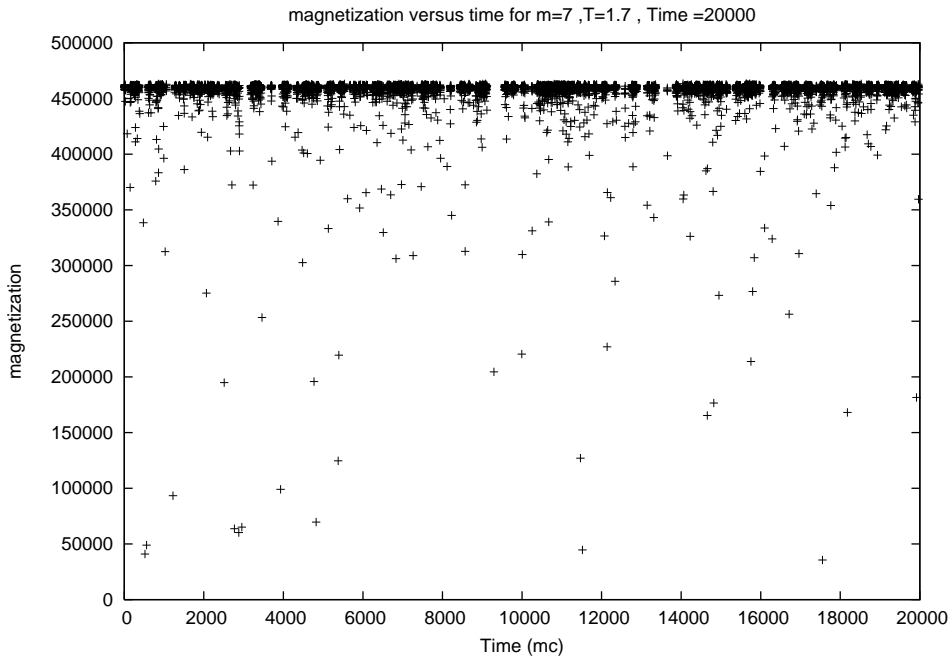}
\end{center}
\caption{
Magnetization versus MCS per spin,  for $N=500000$, time up to 20000,  $m=7$,
for temperature $T = 1.7$.
}
\end{figure}

We determine the time $\tau$ after which the magnetization has decayed to 
3/4 of its initial value (here 375,000) for the first time, and then take 
the median value of our nine samples.
So we get different values of $\tau_1$  for different
temperatures. Then we plot them double-logarithmically versus temperature 
in Fig.2 and as $1/\ln\tau$ versus temperature in Fig. 3.

\begin{figure}[hbt]
\begin{center}
\includegraphics[angle=-90,scale=0.5]{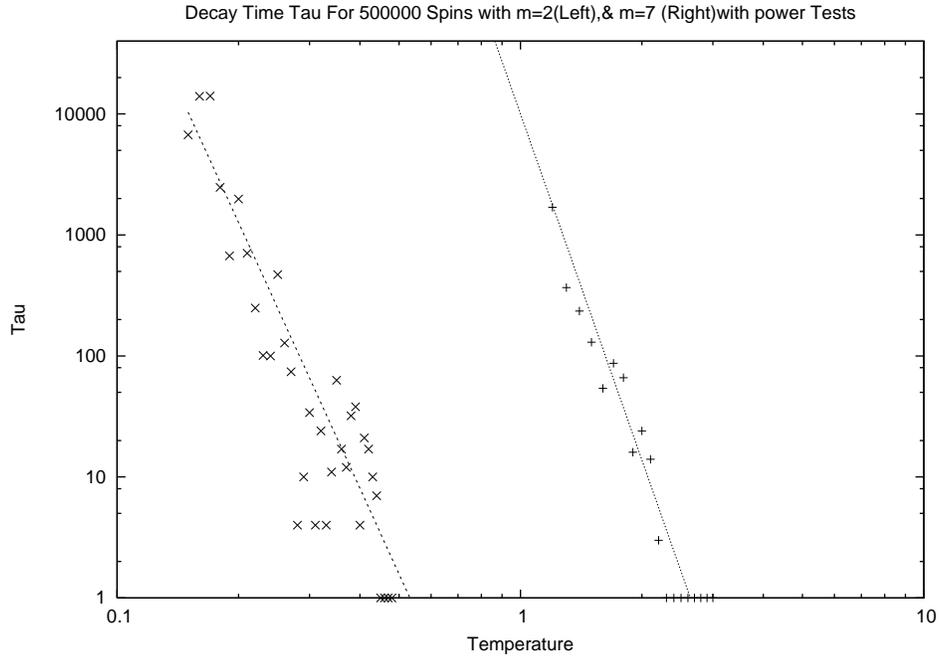}
\end{center}
\caption{
$\tau$ versus  temperature for $N=500000$, time up to 20000, $m=2$ (left)
and 7 (right). Each symbol is the median of nine samples. The straight lines
have different slopes and thus correspond to power laws with different 
exponents.
}
\end{figure}

\begin{figure}[hbt]
\begin{center}
\includegraphics[angle=-90,scale=0.5]{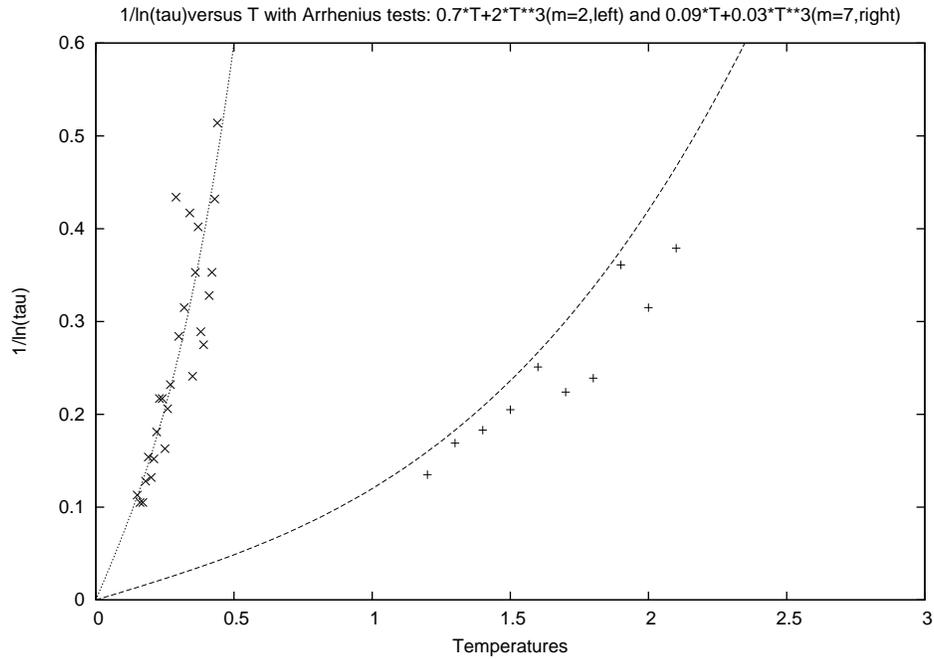}
\end{center}
\caption{
$1/\ln(\tau)$ versus  temperature same data as in Fig. 2
The curves are parabolas correponding to an asymptotic
Arrhenius law $\tau \propto \exp({\rm const}(m)/T)$. 
}
\end{figure}

Since the power-law test of Fig.2 gives two different exponents, we prefer 
the fits of Fig.3 to an Arrhenius law as given in the headline of the figure.

\bigskip
{\bf Conclusion}:

We see that our results agree with the modified Arrhenius law:

$$1/\ln(\tau) = {\rm const}(m) \cdot T + O(T^3)$$ 
meaning that for each positive temperature there is a finite  relaxation time
after which the initial magnetization decays towards zero: Similar to the 
one-dimensional Ising model there is no ferromagnetism on this directed
Barabasi-Albert network.

\bigskip
{\bf References}:

\parindent 0pt

[1] David P. Landau, Kurt Binder, A guide to Monte Carlo simulation in 
statistical physics; Cambridge University Press (2002).

[2] R. Albert and A.L. Barabasi, Rev. Mod. Phys. 74, 47 (2002).

[3] A. Aleksiejuk, J. A. Holyst, D. Stauffer, Physica A 310, 260 (2002);
J.O. Indekeu,  Physica A 333, 461 (2004).

[4] D. Stauffer and H. Meyer-Ortmanns, Int. J. Mod. Phys.C 15, 241 (2004).
\end{document}